\documentclass[aps,prd,superscriptaddress,twoside,twocolumn,nofootinbib,%
10pt,showpacs,floatfix]{revtex4-2}

\usepackage{amsmath,amssymb}
\usepackage{graphicx,bm}
\usepackage{ulem}
\usepackage{slashed}
\usepackage{dcolumn}
\usepackage{epsfig}
\usepackage{multirow}

\allowdisplaybreaks

\begin{document}


\title{Role of hidden-color components in the tetraquark mixing model}


\author{Hungchong Kim}%
\email{bkhc5264@korea.ac.kr}
\affiliation{Center for Extreme Nuclear Matters, Korea University, Seoul 02841, Korea}

\author{K. S. Kim}%
\email{kyungsik@kau.ac.kr}
\affiliation{School of Liberal Arts and Science, Korea Aerospace University, Goyang, 412-791, Korea}

\date{\today}


\begin{abstract}
Multiquarks can have two-hadron components and hidden-color components in their wave functions.
The presence of two-hadron components in multiquarks introduces a potential source of confusion,
particularly with respect to their resemblance to hadronic molecules.
On the other hand, hidden-color components are essential for distinguishing between multiquarks and hadronic molecules.
In this work, we study the hidden-color components in the wave functions of the tetraquark mixing model, a model that has been proposed as a suitable
framework for describing the properties of two nonets in the $J^P=0^+$ channel: the light nonet
[$a_0 (980)$, $K_0^* (700)$, $f_0 (500)$, $f_0 (980)$] and the heavy nonet [$a_0 (1450)$, $K_0^* (1430)$, $f_0 (1370)$, $f_0 (1500)$].
Our analysis reveals a substantial presence of hidden-color components within the tetraquark wave functions.
To elucidate the impact of hidden-color components on physical quantities, we conduct computations of the hyperfine masses, $\langle V_{CS}\rangle$,  for the two nonets,
considering scenarios involving only the two-meson components and those incorporating the hidden-color components.
We demonstrate that the hidden-color components constitute an important part of the hyperfine masses,
such that the mass difference formula, $\Delta M\approx \Delta \langle V_{CS}\rangle$, which has been successful for the two nonets,
cannot be achieved without the hidden-color contributions. This can provide another evidence supporting the tetraquark nature of the two nonets.
\end{abstract}

\maketitle

\section{Introduction}

Multiquarks are a new type of hadron composed of four or more constituent quarks.
These include tetraquarks, pentaquarks, hexaquarks and so on. Multiquarks are different from
normal hadrons that are composed of two or three constituent quarks.
Since there is no particular reason for multiquarks not to exist, they have been long anticipated.
Recently, promising candidates for multiquarks have been reported especially in the heavy quark system, including
$\chi_{c1} (3872)$, $X^{\pm} (4020)$, $\chi_{c1} (4140)$,
$Z_c (3900)$~\cite{Belle03, BESIII:2013ouc, LHCb:2016axx, Xiao:2013iha}, $T_{cc}^+ (3875)$~\cite{LHCb:2021auc,LHCb:2021vvq}, $P_c (4312)$, $P_c (4440)$ and $P_c(4457)$~\cite{LHCb:2015yax,LHCb:2019kea}.
In the light-quark system, there is a well-known candidate for tetraquarks, namely the light nonet
consisting of $a_0 (980)$, $K_0^* (700)$, $f_0 (500)$, and $f_0 (980)$~\cite{Jaffe77a, Jaffe77b, Jaffe04}.
And, the heavy nonet composed of $a_0 (1450)$, $K_0^* (1430)$, $f_0 (1370)$, and $f_0 (1500)$ is also
expected to be tetraquarks generated by the tetraquark mixing model~\cite{Kim:2016dfq, Kim:2017yvd, Kim:2018zob,Lee:2019bwi,Kim:2017yur,Kim:2022qfj,Kim:2023bac,Kim:2019mbc,Kim:2023tph}.

Currently, one of the main problems in clarifying these candidates as multiquarks is that they can also be described as
composite systems of hadrons, which are often referred to as hadronic molecules~\cite{Guo:2017jvc}.
In this description, the candidates can be treated as states of two color-singlets,
such as meson-meson bound systems, meson-baryon systems, or the states that are dynamically generated from two hadrons~\cite{Janssen:1994wn,Weinstein:1990gu,Branz:2007xp,Branz:2008ha}.
For example, the $\chi_{c1}$(3872) could be a tetraquark with the flavor structure
of $cq\bar{c}\bar{q}~ (q=u,d)$~\cite{Maiani:2004vq, Kim:2016tys}, or it could be a meson molecular state
of $D\bar{D}^*$~\cite{Tornqvist:2004qy, Tornqvist:1993ng}. The $T_{cc}^+ (3875)$ might be a double-charmed tetraquark, $cc\bar{q}\bar{q}$,
or it could be a composite state of $DD^*$~\cite{Ohkoda:2012hv,Sakai:2023syt}. The pentaquark candidates such as the $P_c (4312)$, $P_c (4440)$, $P_c$(4457) resonances
can also be described as the hadronic molecules, such as
$\Sigma_c \bar{D}$ ($J^P=1/2^-$), $\Sigma_c \bar{D}^*$ ($J^P=3/2^-$), $\Sigma_c \bar{D}^*$ ($J^P=1/2^-$), respectively~\cite{Du:2019pij,Xiao:2019mvs}.
The $d^*(2380)$ resonance reported in Ref.~\cite{WASA-at-COSY:2011bjg} could be a hexaquark state~\cite{Kim:2020rwn} or
a $\Delta\Delta$ molecular state as predicted by Dyson and Xuong~\cite{Dyson:1964xwa}.
There is similar confusion in identifying the two nonets in the light quark system, the light nonet and the heavy nonet, as tetraquarks.
Therefore, to confirm multiquarks, it is necessary to develop specific strategies for distinguishing them from hadronic molecules.

To do this, we examine the key differences between multiquarks and hadronic molecules.
In color space, hadronic molecules are composed of two color-singlet objects, so they have two-hadron components only.
Multiquarks also have two-hadron components but they can additionally possess hidden-color components.
To elaborate this point, let us consider the tetraquark in the diquark-antidiquark model~\cite{Jaffe77a, Jaffe77b, Jaffe04},
which we denote as $q_1q_2\bar{q}_3\bar{q}_4$.
By construction, this tetraquark forms a color-singlet state by combining the diquark $q_1q_2$ in $\bar{\bm{3}}_c$ and
the antidiquark $\bar{q}_3\bar{q}_4$ in $\bm{3}_c$.
Now, if this is rearranged into two quark-antiquark pairs, like $(q_1\bar{q}_3)(q_2\bar{q}_4)$,
one can clearly see that this tetraquark has two-meson components consisting of two color-singlets, $(q_1\bar{q}_3)_{\bm{1_c}} (q_2\bar{q}_4)_{\bm{1_c}}$,
and hidden-color components of $[(q_1\bar{q}_3)_{\bm{8_c}} (q_2\bar{q}_4)_{\bm{8_c}}]_{\bm{1_c}}$, representing
a color-singlet state constructed from two color-octets.
Therefore, the hidden-color components,
which in this case refer to the internal structure represented by mesonic ($\in \bm{8}_c$) subcomponents,
are unique to tetraquarks and should not appear in hadronic molecules.
Note that the decomposition in color space is universal for other multiquarks as well~\cite{Sazdjian:2022fyq}.

With this difference in mind, two strategies can be adapted to distinguish multiquarks and hadronic molecules.
The first strategy is to examine the two-hadron components, either in a multiquark framework or through a hadronic molecular perspective.
The second strategy is demonstrating the contribution of hidden-color components that exist only in the multiquark wave function.
These two strategies can be used to solidify the tetraquark mixing model that has been proposed as a relevant framework for the above two nonets,
namely the light nonet and the heavy nonet.

According to the tetraquark mixing model, the two nonets are expected to be tetraquarks formed by mixing two types of tetraquarks~\cite{Kim:2016dfq, Kim:2017yvd, Kim:2018zob,Lee:2019bwi,Kim:2017yur,Kim:2022qfj,Kim:2023bac,Kim:2019mbc,Kim:2023tph}. Alternatively,
some members of the two nonets can be described as hadronic molecules~\cite{Ahmed:2020kmp,Molina:2008jw}.
The two models can be distinguished by the strength of the coupling between the two nonet and two meson components.
Specifically, the tetraquark mixing model predicts that the coupling strengths with two pseudoscalar mesons are enhanced in the light nonet but suppressed in the heavy nonet~\cite{Kim:2017yur,Kim:2022qfj,Kim:2023bac}.
This prediction arises from the fact that the two-meson components of the tetraquark wave functions, through the mixing formulas, add up in the light nonet but partially cancel each other in the heavy nonet.
In contrast, reproducing this result in the framework of meson molecules is currently unclear due to the difficulty in constructing two realistic nonets
from meson molecules, which consequently prevents the establishment of a mixing mechanism leading to the discriminating couplings of the two nonets~\cite{Kim:2023tph}.
Interestingly, the prediction from the tetraquark model is well supported by the experimental partial decay widths of both nonets~\cite{Kim:2017yur,Kim:2022qfj,Kim:2023bac}.

The two-meson components can lead to different decay modes depending on whether the two nonets are described by the tetraquark mixing model or
by meson molecules.
In the tetraquark mixing model, possible decay modes are determined by the fall-apart mechanism where quark-antiquark pairs
in the wave functions simply fall apart into two mesons.
In the framework of meson molecules, however,
the decay modes are dictated by the possible two-meson modes allowed by the SU$_f$(3) symmetry when constructing two-meson states.
In fact, the decay modes predicted by the tetraquark
mixing model align more closely with experimental data~\cite{Kim:2023tph}.

The second strategy, which utilizes hidden-color components, can also be used to identify the two nonets
as tetraquarks. As mentioned, these hidden-color components exist only in tetraquarks.
This feature provides a direct method of identifying tetraquarks if we
can demonstrate that these hidden-color components indeed affect certain physical quantities.
In this context, we investigate how these hidden-color components contribute to hyperfine mass,
defined by the expectation value of the color-spin interaction.
The hyperfine mass, which actually refers to contributions from hyperfine interactions
between any two quarks in the bound states, constitutes an
important part of the physical mass and demonstrates a successful aspect of the tetraquark mixing model for the two nonets.
When calculated from the tetraquark mixing model, the hyperfine mass of the light nonet is found to have a large negative value,
providing a qualitative explanation for why the light nonet can have masses less than 1 GeV.
On the other hand, the hyperfine mass for the heavy nonet is found to be small so
it can provide a qualitative rationale for why
the masses of the heavy nonet are not significantly different from four times the constituent quark mass.
More importantly, the mass difference between the two nonets is well accounted for by the hyperfine mass splitting.
Therefore, the hyperfine mass can be considered a physical quantity to verify the influence of hidden-color components.

In this study, we investigate the role of hidden-color components on the hyperfine masses of two nonets within the tetraquark mixing model.
In the tetraquark mixing model, the wave functions for the two nonets are represented by a mixture of two tetraquark types.
In our approach, we begin with separating the two tetraquark types into two-meson components and hidden-color components.
Through the mixing formulas, the resulting two types can be used to further decompose the wave functions in the tetraquark mixing model.
Using the decomposed form of the wave functions, we calculate the physical hyperfine mass with and without hidden-color components.
This allows us to examine how important hidden-color components are in generating physical quantities such as the hyperfine mass.

This paper is organized as follows. In Sec.~\ref{tetraquark types}, we introduce two tetraquark types in the diquark-antidiquark
form that have been used to formulate the tetraquark mixing model. The explicit expressions for the two tetraquark types
will be presented separately in flavor, spin, and color space.
In Sec.~\ref{decomposition}, we rearrange the two tetraquark types into two quark-antiquark pairs, decomposing the two types into two-meson
components and hidden-color components.
In Sec.~\ref{tetraquark minxing}, this type of decomposition is further applied to the physical wave functions from the tetraquark mixing model using
the results in Sec.~\ref{decomposition}.
We then investigate the role of hidden-color components in the hyperfine mass in Sec.~\ref{hyperfine}.

\section{Two tetraquark types}
\label{tetraquark types}

The tetraquark mixing model~\cite{Kim:2016dfq,Kim:2017yvd,Kim:2018zob,Lee:2019bwi,Kim:2017yur,Kim:2022qfj,Kim:2023bac,Kim:2019mbc,Kim:2023tph} utilizes two types of tetraquarks, consisting of diquarks and antidiquarks.
In this section, we explain the two types of tetraquarks written separately in spin, color, and flavor spaces as:
\begin{eqnarray}
&&~~~~~~~~~~~~~~~\underline{\text{spin}}~~~~~\underline{\text{color}}~~~~~~~~~\underline{\text{flavor}}\nonumber \\
&&|\text{Type1} \rangle = | 000 \rangle \otimes |\bm{1}_c \bar{\bm{3}}_c \bm{3}_c\rangle \otimes |\bm{9}_f \bar{\bm{3}}_f \bm{3}_f\rangle \label{type1}\ ,\\
&&|\text{Type2} \rangle = | 011 \rangle \otimes |\bm{1}_c \bm{6}_c \bar{\bm{6}}_c\rangle \otimes |\bm{9}_f \bar{\bm{3}}_f \bm{3}_f\rangle \label{type2}\ .
\end{eqnarray}
In these expressions~\footnote{In our previous studies~\cite{Kim:2016dfq, Kim:2017yvd,Kim:2018zob,Lee:2019bwi,Kim:2017yur,Kim:2022qfj,Kim:2023bac,Kim:2019mbc,Kim:2023tph}, $|\text{Type1} \rangle$ and $|\text{Type2} \rangle$ were simply denoted by $|000 \rangle$ and $|011 \rangle$, respectively.},
the first number represents the state of the tetraquark, the second number represents the state of the diquark, and the third number represents the state of the antidiquark.
For example, in color space, $|\bm{1}_c \bm{6}_c \bar{\bm{6}}_c\rangle$ represents a color-singlet tetraquark constructed by the $\bm{6}_c$ diquarks and the $\bar{\bm{6}}_c$ antidiquarks.
Therefore, the tetraquark type, $|\text{Type1}\rangle$, denotes the spin-0 tetraquark constructed by combining a spin-0
diquark of the color and flavor structure, ($\bar{\bm{3}}_c, \bar{\bm{3}}_f$), along with a spin-0 antidiquark of ($\bm{3}_c, \bm{3}_f$).
The other tetraquark type, $|\text{Type2}\rangle$, also denotes the spin-0 tetraquark, but is constructed by
combining a spin-1 diquark of ($\bm{6}_c, \bar{\bm{3}}_f$) with a spin-1 antidiquark of ($\bar{\bm{6}}_c, \bm{3}_f$).

A few remarks are in order.
In this approach, the spin-0 diquark is utilized in constructing $|\text{Type1}\rangle$ because it is the most compact diquark among all the
possible diquarks~\cite{Jaffe77a, Jaffe77b, Jaffe04}.
But for $|\text{Type2}\rangle$, the spin-1 diquark is employed because it is the second most compact among all the possible diquarks.
Both types, $| \text{Type1} \rangle$ and $|\text{Type2} \rangle$,
share a common flavor structure of $\bm{9}_f$, which is formed by $\bar{\bm{3}}_f\otimes \bm{3}_f =\bm{9}_f(=\bm{8}_f\oplus \bm{1}_f$).
This flavor structure gives rise to an ``inverted mass ordering'',
the mass hierarchy observed in the light nonet and, to a lesser extent, the heavy nonet~\footnote{The inverted mass ordering,
referring to the sequence in the light nonet, $M[a_0(980)] > M[K^*_0(700)] > M[f_0(500)]$,
is opposite to the expected mass ordering of $M[a_0(980)]<M[K^*_0(700)]<M[f_0(500)]$ in a two-quark picture.
The inverted mass ordering has been proposed as crucial evidence supporting that the light
nonet members are tetraquarks~\cite{Jaffe77a,Jaffe77b,Jaffe04}.}.
Therefore, the two types can be used to describe the two nonets within the mixing mechanism to be discussed below.
Another remark is that the tetraquark under consideration is in the ground state, meaning all of its constituent quarks
are in the $S$-wave. Consequently, the spatial wave functions are symmetric under the exchange of any pair of quarks~\cite{Silvestre-Brac:1992kaa}.
Due to this symmetry requirement, diquarks and antidiquarks in $| \text{Type1} \rangle$ and $|\text{Type2} \rangle$ must be
antisymmetric in the combined space of spin-color-flavor.  The spin, color, and flavor structure
in Eqs.~(\ref{type1}) and (\ref{type2}) ensures that diquarks and antidiquarks satisfy this antisymmetric condition.
Below, we present the mathematical structure of the two types in flavor, spin, and color space.

\subsection{Flavor part}
\label{wflavor}

In flavor space, the two tetraquark types, $|\text{Type1}\rangle$ and $|\text{Type2}\rangle$, separately form a nonet.
Each nonet can potentially represent nine states in the $J^P=0^+$ channel, analogous to the light nonet or the heavy nonet.
By adopting a tensor notation~\footnote{Ref.~\cite{Oh:2004gz} might be useful for technical details in using the tensor notation.},
one can express all the members in term of diquarks ($T^i$) and antidiquarks ($\bar{T}_i$) as follows:
\begin{eqnarray}
[{\bf 8}_f]^i_{j} &=& T^{i}\bar{T}_{j}-\frac{1}{3} \delta^{i}_{j}~ T^{m}\bar{T}_{m}\label{flavor octet}\ ,\\
{\bf 1}_f &=& \frac{1}{\sqrt{3}}T^{m}\bar{T}_{m}\label{flavor singlet}\ .
\end{eqnarray}
Here $T^i$ ($\bar{T}_i$) is an antisymmetric combination of quarks (antiquarks) defined as:
\begin{eqnarray}
T^i &=&\frac{1}{\sqrt{2}}\epsilon^{ijk}q^j q^k\equiv [q^j q^k]\ ,\nonumber\\
\bar{T}_i &=& \frac{1}{\sqrt{2}}\epsilon_{ijk}\bar{q}_j \bar{q}_k \equiv [{\bar q}_j {\bar q}_k] \ .
\end{eqnarray}
In terms of explicit quark flavors ($q^i=u,d,s$), all the members of Eqs.~(\ref{flavor octet}),(\ref{flavor singlet}) are expressed as
\begin{eqnarray}
&&[\bm{8}_f]^3_1 =[ud][\bar{d}\bar{s}]\label{o31} ,\\
&&[\bm{8}_f]^3_2 =[ud][\bar{s}\bar{u}]\label{o32} ,\\
&&[\bm{8}_f]^2_1 =[su][\bar{d}\bar{s}]\label{o21} ,\\
\frac{1}{\sqrt{2}}\{[\bm{8}_f]^1_1 &&- [\bm{8}_f]^2_2\} =\frac{1}{\sqrt{2}}\left\{[ds][\bar{d}\bar{s}]-[su][\bar{s}\bar{u}]\right\}\label{o1122} ,\\
&&[\bm{8}_f]^1_2 = [ds][\bar{s}\bar{u}]\label{o12}\ , \\
\sqrt{\frac{3}{2}}[\bm{8}_f]^3_3&& = \frac{1}{\sqrt{6}}\left\{2[ud][\bar{u}\bar{d}]-[ds][\bar{d}\bar{s}]-[su][\bar{s}\bar{u}] \right\} \label{o33} ,\\
&&[\bm{8}_f]^2_3 =[su][\bar{u}\bar{d}]\label{o23} ,\\
&&[\bm{8}_f]^1_3 =[ds][\bar{u}\bar{d}]\label{o13} ,\\
\bm{1}_f&&=\frac{1}{\sqrt{3}}\left\{[ds][\bar{d}\bar{s}]+[su][\bar{s}\bar{u}]+[ud][\bar{u}\bar{d}]\right\}\label{singlet} .
\end{eqnarray}

\subsection{Spin part}
\label{wspin}

The spin parts, $|000\rangle$ and $|011\rangle$, in Eqs.~(\ref{type1}),(\ref{type2}), can also be expressed in terms of the spin states of
diquarks and antidiquarks.
To do this, we first label the diquark (antidiquark) as $q_1q_2$ ($\bar{q}_3\bar{q}_4$).
Then, in $|000\rangle$, the diquark has a spin of $J_{12}=0$ and its projection
$M_{12}=0$, so the spin state of the diquark is denoted by $|J_{12}, M_{12}\rangle=|0,0\rangle_{12}$.
Similarly, the spin state of the antidiquark is denoted by $|J_{34}, M_{34}\rangle=|0,0\rangle_{34}$.
Thus, the spin-0 tetraquark from these can be expressed as:
\begin{eqnarray}
|000 \rangle =|0,0\rangle_{12}|0,0\rangle_{34}\label{000a} .
\end{eqnarray}
The other spin configuration, $|011\rangle$, consists of the diquark with spin $J_{12}=1$ and the antidiquark with $J_{34}=1$.
The spin-0 tetraquark state constructed from these can be obtained readily as:
\begin{eqnarray}
&&|011\rangle = \frac{1}{\sqrt{3}} \Big \{ |1,1\rangle_{12} |1,-1\rangle_{34} - |1,0\rangle_{12} |1,0\rangle_{34} \nonumber \\
&&~~~~~~~~~+ |1,-1\rangle_{12} |1,1\rangle_{34}] \Big \}\label{011a}\ .
\end{eqnarray}
Note that Eqs.~(\ref{000a}), (\ref{011a}) are normalized to unity, $\langle 000|000\rangle=1$, $\langle 011|011\rangle=1$.

\subsection{Color part}
\label{wcolor}

One can write the color parts, $| \bm{1}_c \bar{\bm{3}}_c \bm{3}_c \rangle$ and $|\bm{1}_c \bm{6}_c \bar{\bm{6}}_c\rangle$,
in Eqs.~(\ref{type1}) and (\ref{type2}), respectively, in terms of the color states of diquarks and antidiquarks.
The color configuration, $|\bm{1}_c \bar{\bm{3}}_c \bm{3}_c \rangle$ in $|\text{Type1}\rangle$,
represents a color-singlet tetraquark constructed by combining diquarks in $\bar{\bm{3}}_c$ and antidiquarks in $\bm{3}_c$,
i.e.,  $\bar{\bm{3}}_c\otimes\bm{3}_c\rightarrow \bm{1}_c$.
In tensor notation, this color configuration can be written as
\begin{eqnarray}
&&| \bm{1}_c \bar{\bm{3}}_c \bm{3}_c \rangle =\frac{1}{\sqrt{12}}  \epsilon_{abd} \epsilon^{aef}
\big ( q^b_1 q^d_2 \big )
\big ( \bar{q}_{3e} \bar{q}_{4f} \big )\label{color1}\ .
\end{eqnarray}
Here the indices, $a,b,d,e,f$, represent colors.
For $|\text{Type2}\rangle$, its color configuration $|\bm{1}_c \bm{6}_c \bar{\bm{6}}_c\rangle$ denotes a color-singlet
tetraquark constructed by combining diquarks in $\bm{6}_c$ and antidiquarks in $\bar{\bm{6}}_c$, i.e., $\bm{6}_c \otimes \bar{\bm{6}}_c \rightarrow \bm{1}_c$.
In tensor notation, it can be written as
\begin{eqnarray}
&&|\bm{1}_c \bm{6}_c \bar{\bm{6}}_c\rangle= \frac{1}{\sqrt{96}} \big( q^a_1 q^b_2+q^b_1 q^a_2 \big )
\big (\bar{q}_{3a} \bar{q}_{4b}+\bar{q}_{3b} \bar{q}_{4a}\big )\label{color2}\ .
\end{eqnarray}

\section{Decomposition of the two tetraquark types}
\label{decomposition}

The two types introduced above, $|\text{Type1}\rangle$ and $|\text{Type2}\rangle$, represent two possible tetraquarks constructed as
$(q_1q_2)(\bar{q}_3\bar{q}_4)$ in the diquark-antidiquark configuration.
To investigate the influence of hidden-color components, we need to rearrange the tetraquark wave functions into the quark-antiquark pairs, $(q_1\bar{q}_3)(q_2\bar{q}_4)$,
which decomposes each wave function into a two-meson component and a hidden-color component.
This enables us to assess the significance of the hidden-color element.
In later discussion, we will examine the importance of the hidden-color component in the hyperfine mass, which mostly depends on
the color and spin parts of the wave functions. Since the flavor part can be trivially implemented in the hyperfine mass,
it suffices to consider the spin and color parts of the wave functions in this rearrangement.

In the rearrangement in spin space, two quark-antiquark pairs, $(q_1\bar{q}_3)$ and $(q_2\bar{q}_4)$,
are either both in the spin-0 ($J^P=0^-$) state denoted by $PP$ or both in the spin-1 ($J^P=1^-$) state denoted by $VV$.
Specifically, the spin parts of Eqs.~(\ref{000a}) and (\ref{011a}) can be decomposed as
\begin{eqnarray}
&&|000\rangle =\frac{1}{2}PP+\frac{\sqrt{3}}{2}VV\label{spin1} ,\\
&&|011\rangle = \frac{\sqrt{3}}{2}PP-\frac{1}{2}VV\label{spin2}\ ,
\end{eqnarray}
in the rearrangement. Here $PP$ and $VV$ are defined as
\begin{eqnarray}
&&PP =|0,0\rangle_{13} |0,0\rangle_{24}\label{PP} ,\\
&&VV = \frac{1}{\sqrt{3}} \Big \{ |1,1\rangle_{13} |1,-1\rangle_{24} - |1,0\rangle_{13} |1,0\rangle_{24} \nonumber \\
&&~~~~~~~~~+ |1,-1\rangle_{13} |1,1\rangle_{24}] \Big \}\label{VV}\ .
\end{eqnarray}
Again, we notice that $PP$ and $VV$ are orthonormal.
A detailed derivation of these or related discussion can be found in Refs.~\cite{Kim:2018zob,Black:1998wt}.

One can do a similar rearrangement in color space. The color configurations,
expressed in the diquark-antidiquark form [Eqs.~(\ref{color1}) and (\ref{color2})], can be
rearranged into $(q_1q_2)(\bar{q}_3\bar{q}_4)\rightarrow (q_1\bar{q}_3)(q_2\bar{q}_4)$.
To do this, we begin with the fact that any quark-antiquark pair in color space can be written in terms of an octet and
a singlet as,
\begin{eqnarray}
q^a \bar{q}_b =(\bm{8}_c)^a_b+\frac{1}{\sqrt{3}}\delta^a_b\bm{1}_c\label{qqbar}\ ,
\end{eqnarray}
where
\begin{eqnarray}
(\bm{8}_c)^a_b &=& q^a\bar{q}_{b}- \frac{1}{3} \delta^a_b q^d\bar{q}_d\ ,\\
\bm{1}_c &=& \frac{1}{\sqrt{3}} q^d\bar{q}_d\ .
\end{eqnarray}
Note that, the color-singlet is normalized to unity, $\langle \bm{1}_c|\bm{1}_c\rangle=1$, and the color-octet is normalized according to
\begin{eqnarray}
\langle (\bm{8}_c)^a_b|(\bm{8}_c)^{a^\prime}_{b^\prime}\rangle =  \delta^{aa^\prime}\delta_{bb^\prime}-\frac{1}{3}\delta^a_b\delta^{a^\prime}_{b^\prime} .
\end{eqnarray}
Applying Eq.~(\ref{qqbar}) to the quark-antiquark pairs, $q_1\bar{q}_3$ and $q_2\bar{q}_4$, in Eqs.~(\ref{color1}),(\ref{color2}),
one can decompose the color configurations straightforwardly as
\begin{eqnarray}
&&| \bm{1}_c \bar{\bm{3}}_c \bm{3}_c \rangle =\frac{1}{\sqrt{3}} \big [ (\bm{1}_c)_{13} (\bm{1}_c)_{24} \big ] \nonumber \\
&&~~~~~~~~~~~~~~~~ - \sqrt{\frac{2}{3}} \Bigg \{\frac{1}{2\sqrt{2}} \text{Tr}\left [(\bm{8}_c)_{13} (\bm{8}_c)_{24}\right ] \Bigg \}\label{col1} ,\\
&&|\bm{1}_c \bm{6}_c \bar{\bm{6}}_c\rangle =\sqrt{\frac{2}{3}}  \big [ (\bm{1}_c)_{13} (\bm{1}_c)_{24} \big ] \nonumber \\
&&~~~~~~~~~~~~~~~~ + \frac{1}{\sqrt{3}} \Bigg \{\frac{1}{2\sqrt{2}} \text{Tr}\left [(\bm{8}_c)_{13} (\bm{8}_c)_{24}\right ] \Bigg \}\label{col2}\ .
\end{eqnarray}

In this decomposition, the component containing $(\bm{1}_c)_{13} (\bm{1}_c)_{24}$ represents a two-meson state composed of two color-singlets,
$q_1\bar{q}_3 \in\bm{1}_c$ and $q_2\bar{q}_4\in\bm{1}_c$.
The presence of this component can potentially lead to confusion, as it alone resembles meson molecules.
Tetraquarks can decay into two mesons through this component if the decay modes are energetically allowed.
The other component involving $\frac{1}{2\sqrt{2}} \text{Tr}\left [(\bm{8}_c)_{13} (\bm{8}_c)_{24}\right ]$ represents a
color-singlet state constructed from $q_1\bar{q}_3 \in\bm{8}_c$ and $q_2\bar{q}_4\in\bm{8}_c$.
This is the hidden-color component that can distinguish tetraquarks from two-meson molecules.
Note, $(\bm{1}_c)_{13} (\bm{1}_c)_{24}$
and $\frac{1}{2\sqrt{2}} \text{Tr}\left [(\bm{8}_c)_{13} (\bm{8}_c)_{24}\right ]$
are orthogonal and normalized to unit magnitude separately.
From this decomposition, it is evident that the contribution of the hidden-color component is significant in both configurations.
Specifically, it constitutes two-thirds in $| \bm{1}_c \bar{\bm{3}}_c \bm{3}_c \rangle$ and one-third in $|\bm{1}_c \bm{6}_c \bar{\bm{6}}_c\rangle$.

One can make an alternative rearrangement by applying Eq.~(\ref{qqbar}) to different quark-antiquark pairs, $q_2\bar{q}_3$ and $q_1\bar{q}_4$,
when decomposing Eqs.~(\ref{color1}),(\ref{color2}).
This step also generates the expressions like Eqs.~(\ref{col1}),(\ref{col2}) but with the trivial changes in labeling, $(13)\rightarrow (23)$ and $(24)\rightarrow (14)$.
Consequently, recombining into $q_1\bar{q}_4$ and $q_2\bar{q}_3$ pairs does not lead to different physical results in the end.

One may argue that the terms containing $\frac{1}{2\sqrt{2}} \text{Tr}\left [(\bm{8}_c)_{13} (\bm{8}_c)_{24}\right ]$ in Eqs.~(\ref{col1}),(\ref{col2})
are not a true hidden-color
component because they can produce two-meson components when further rearranged in terms of $q_2\bar{q}_3$ and $q_1\bar{q}_4$.
However, the second rearrangement of the terms should not be performed without entailing an identical
rearrangement in the terms containing\footnote{
Physically, quark confinement does not allow the second rearrangement on $(\bm{1}_c)_{13} (\bm{1}_c)_{24}$
in terms of $q_2\bar{q}_3$ and $q_1\bar{q}_4$.
Our discussion here is purely mathematical, aiming to prove that the second rearrangement is not
permissible when applied solely to $\frac{1}{2\sqrt{2}} \text{Tr}\left [(\bm{8}_c)_{13} (\bm{8}_c)_{24}\right ]$. Thus, the physical constraint
is not needed to be imposed in this derivation.
}
$(\bm{1}_c)_{13} (\bm{1}_c)_{24}$. This is because two rearrangements, $(q_1\bar{q}_3)(q_2\bar{q}_4)$ and $(q_1\bar{q}_4)(q_2\bar{q}_3)$, cannot be separately performed in the original
configuration in $| \bm{1}_c \bar{\bm{3}}_c \bm{3}_c \rangle$
or $|\bm{1}_c \bm{6}_c \bar{\bm{6}}_c\rangle$.
So, if one makes the same rearrangement into $q_2\bar{q}_3$ and $q_1\bar{q}_4$ also in the $(\bm{1}_c)_{13} (\bm{1}_c)_{24}$ terms,
the resulting decomposition would be the same as the one obtained by rearranging the original configuration,
$|\bm{1}_c \bar{\bm{3}}_c \bm{3}_c \rangle$ or $|\bm{1}_c \bm{6}_c \bar{\bm{6}}_c\rangle$, with respect to $q_2\bar{q}_3$ and $q_1\bar{q}_4$.

Later, we will concentrate on the hidden-color contributions to the hyperfine mass that depends on both the color and spin parts of the wave functions.
For this purpose, we combine the color and spin parts by multiplying Eqs.~(\ref{spin1}), (\ref{spin2}) by Eqs.~(\ref{col1}),(\ref{col2}) respectively.
We denote these color-spin parts as $|\text{CS1}\rangle$, $|\text{CS2}\rangle$ whose expressions are as follow:
\begin{eqnarray}
&&|\text{CS1}\rangle =\Big[\frac{1}{2}PP+\frac{\sqrt{3}}{2}VV \Big]
\otimes \Bigg \{\frac{1}{\sqrt{3}} \Big [ (\bm{1}_c)_{13} (\bm{1}_c)_{24} \Big ]\nonumber \\
&&~~~~~~~~~~- \sqrt{\frac{2}{3}} \Big [\frac{1}{2\sqrt{2}} \text{Tr}\left [(\bm{8}_c)_{13} (\bm{8}_c)_{24}\right ] \Big ] \Bigg \}\label{cs1} ,\\
&&|\text{CS2}\rangle =\Big [\frac{\sqrt{3}}{2}PP-\frac{1}{2}VV\Big]
\otimes\Bigg \{ \sqrt{\frac{2}{3}}  \Big [ (\bm{1}_c)_{13} (\bm{1}_c)_{24} \Big ]\nonumber \\
&&~~~~~~~~~~+ \frac{1}{\sqrt{3}} \Big [\frac{1}{2\sqrt{2}} \text{Tr}\left [(\bm{8}_c)_{13} (\bm{8}_c)_{24}\right ] \Big ] \Bigg \}\label{cs2}\ .
\end{eqnarray}
We emphasize once more that $|\text{CS1}\rangle$ and $|\text{CS2}\rangle$ are clearly divided into two parts:
the two-meson (TM) part containing $ (\bm{1}_c)_{13} (\bm{1}_c)_{24}$ and the hidden-color part (HC)
involving $\frac{1}{2\sqrt{2}} \text{Tr}\left [(\bm{8}_c)_{13} (\bm{8}_c)_{24}\right ]$.
Because of this decomposition, we can partition the hyperfine mass into three segments,
which we later denote as TM$\cdot$TM, TM$\cdot$HC, and HC$\cdot$HC.

Note, Eqs.~(\ref{cs1}) and (\ref{cs2}) are obtained from the spin-color parts,
$| 000 \rangle \otimes |\bm{1}_c \bar{\bm{3}}_c \bm{3}_c\rangle$ and
$| 011 \rangle \otimes |\bm{1}_c \bm{6}_c \bar{\bm{6}}_c\rangle$
in Eqs.~(\ref{type1}) and (\ref{type2}), by rearranging $(q_1q_2)(\bar{q}_3\bar{q}_4)$ to $(q_1\bar{q}_3)(q_2\bar{q}_4)$.
Thus, Eqs.~(\ref{cs1}) and (\ref{cs2}) uphold the inherent symmetry
imposed on $| 000 \rangle \otimes |\bm{1}_c \bar{\bm{3}}_c \bm{3}_c\rangle$ and
$| 011 \rangle \otimes |\bm{1}_c \bm{6}_c \bar{\bm{6}}_c\rangle$.
Later, Eqs.~(\ref{cs1}) and (\ref{cs2}) will be combined with the
antisymmetric flavor part to calculate the physical hyperfine mass.  Therefore, the antisymmetric constraints
imposed on Eqs.~(\ref{type1}) and (\ref{type2}) should be maintained regardless of the rearrangement performed in this section.

\section{Decomposition of the tetraquark mixing wave functions}
\label{tetraquark minxing}

The two tetraquark types, $| \text{Type1} \rangle$, $|\text{Type2} \rangle$ in Eqs.~(\ref{type1}) and (\ref{type2}),
share a common flavor structure of $\bm{9}_f$.
However, as shown above, the two types have different structures in color and spin.
An important observation is that the two tetraquark types are strongly mixed through the color-spin
interaction~\cite{Kim:2016dfq, Kim:2017yvd, Silvestre-Brac:1992kaa}~\footnote{
In coordinate space, this potential is often taken to be in the form of $\delta(\bm{r}_{ij})$
so that the symmetric spatial wave functions assumed in this work do not affect the results on hyperfine mass.},
\begin{eqnarray}
V_{CS} &&= \sum_{i < j} \frac{v_0}{m_i^{} m_j^{}} \lambda_i \cdot \lambda_j J_i\cdot J_j \label{VCS}\ .
\end{eqnarray}
Here, $\lambda_i$ denotes the Gell-Mann matrix for the color, $J_i$ the
spin, and $m_i$ the constituent quark mass. Besides $V_{CS}$, the full potential adopted in
Refs.~\cite{Kim:2016dfq, Kim:2017yvd} additionally contains
a color-electric part and a constant shift involving the parameters $v_1$ and $v_2$, respectively.
These additional parts are not considered in this work, as they are not important for the discussion of hyperfine masses and their splitting.

Given that the color-spin interaction is a crucial part of the Hamiltonian in quark systems,
the strong mixing suggests that the two tetraquark types are not eigenstates of the Hamiltonian, indicating that they cannot represent physical states.
Instead, physical states can be identified as combinations of the two tetraquark types that diagonalize the color-spin interaction.
This diagonalization process leads to the physical states which can be collectively expressed by the formulas:
\begin{eqnarray}
|\text{Heavy~nonet} \rangle &=& -\alpha | \text{Type1} \rangle + \beta |\text{Type2} \rangle \label{heavy}\ ,\\
|\text{Light~nonet} \rangle~ &=&~~\beta | \text{Type1} \rangle + \alpha |\text{Type2} \rangle \label{light}\ .
\end{eqnarray}
The eigenstates, $|\text{Heavy~nonet} \rangle$ and $|\text{Light~nonet} \rangle$, are in fact identified by the two physical nonets
in the $J^P=0^+$ channel: the heavy nonet [$a_0 (1450)$, $K_0^* (1430)$, $f_0 (1370)$, $f_0 (1500)$] and
the light nonet [$a_0 (980)$, $K_0^* (700)$, $f_0 (500)$, $f_0 (980)$] in Particle Data Group(PDG)~\cite{PDG22}.
The mixing parameters, $\alpha$ and $\beta$, are also determined from the diagonalization.
Their values are approximately $\alpha \approx \sqrt{2/3}$ and $\beta \approx \sqrt{1/3}$, with slight variations depending on the isospin channel~\cite{Kim:2016dfq,Kim:2017yvd,Kim:2018zob,Kim:2019mbc}~\footnote{In fact, our calculation below uses $\alpha$ and $\beta$
given in Refs.~\cite{Kim:2019mbc,Kim:2017yvd}, which depend slightly on the isospin channel.}.

The model represented by Eqs.~(\ref{heavy}) and (\ref{light}) has been named the tetraquark mixing model, as the two nonets
under consideration are represented by a mixture of two tetraquark types, $|\text{Type1}\rangle$ and $|\text{Type2}\rangle$.
This mixing model appears to be quite successful.
First of all, this model reproduces qualitatively well the masses and mass differences between the two nonets~~\cite{Kim:2016dfq,Kim:2017yvd,Kim:2018zob,Kim:2019mbc}
through the hyperfine mass, $\langle V_{CS}\rangle$.
Second, this model predicts that the coupling strengths of the two nonets with two pseudoscalar mesons
are enhanced in the light nonet but suppressed in the heavy nonet. It is remarkable that
experimental data on the partial decay widths of both nonets strongly support this intriguing prediction~\cite{Kim:2017yur,Kim:2022qfj,Kim:2023bac}.
Third, this tetraquark model is better than meson molecules in explaining the decay modes of the two nonets~\cite{Kim:2023tph}.
Lastly, QCD sum rules constructed for $a_0(980)$ support this model~\cite{Lee:2019bwi}.
All evidence suggests that the two nonets above are
tetraquarks, particularly generated by the mixing formulas, Eqs.~(\ref{heavy}) and (\ref{light}).

As further supporting evidence of the tetraquark nature of the two nonets, we are investigating in this work hidden-color contributions to the hyperfine mass.
Since the hyperfine mass depends on color-spin part of the
wave functions,
we extract the common flavor part in Eqs.~(\ref{heavy}) and (\ref{light}),
and write the mixing formulas solely for the color-spin part as,
\begin{eqnarray}
|\text{HN}\rangle_{CS}~ &=& -\alpha  |\text{CS1} \rangle + \beta |\text{CS2} \rangle \label{hcs}\ ,\\
|\text{LN}\rangle_{CS}~ &=&~~\beta  |\text{CS1} \rangle + \alpha |\text{CS2} \rangle \label{lcs}\ .
\end{eqnarray}

Using the expressions for $|\text{CS1} \rangle$, $|\text{CS2} \rangle$ in Eqs.~(\ref{cs1})(\ref{cs2}),
we can rewrite these mixing wave functions as:
\begin{widetext}
\begin{eqnarray}
&& |\text{HN}\rangle_{CS} =\Big[ \left (-\frac{\alpha}{2\sqrt{3}}+\frac{\beta}{\sqrt{2}} \right ) PP
 - \left (\frac{\alpha}{2}+ \frac{\beta}{\sqrt{6}}\right ) VV \Big ]\otimes \left[(\bm{1}_c)_{13} (\bm{1}_c)_{24}\right] \nonumber \\
&&~~~~~~~~~~~~~~~ + \Big [ \left (\frac{\alpha}{\sqrt{6}}+\frac{\beta}{2} \right ) PP + \left (\frac{\alpha}{\sqrt{2}} - \frac{\beta}{2\sqrt{3}}\right )VV \Big ] \otimes
\Big [\frac{1}{2\sqrt{2}} \text{Tr}\left [(\bm{8}_c)_{13} (\bm{8}_c)_{24}\right ]\Big] \label{HCS1}\ ,\\
&&|\text{LN}\rangle_{CS} =\Big[ \left (\frac{\beta}{2\sqrt{3}}+\frac{\alpha}{\sqrt{2}} \right ) PP
 + \left (\frac{\beta}{2} - \frac{\alpha}{\sqrt{6}}\right ) VV \Big ]\otimes \left[(\bm{1}_c)_{13} (\bm{1}_c)_{24}\right] \nonumber \\
&&~~~~~~~~~~~~~~~ + \Big [ \left (-\frac{\beta}{\sqrt{6}}+\frac{\alpha}{2} \right ) PP - \left (\frac{\beta}{\sqrt{2}} + \frac{\alpha}{2\sqrt{3}}\right )VV \Big ] \otimes
\Big [\frac{1}{2\sqrt{2}} \text{Tr}\left [(\bm{8}_c)_{13} (\bm{8}_c)_{24}\right ]\Big]\label{LCS1}\ .
\end{eqnarray}
\end{widetext}
As evident from Eqs.~(\ref{hcs}),(\ref{lcs}), the state $|\text{LN}\rangle_{CS}$ can be obtained from $|\text{HN}\rangle_{CS}$ simply by replacing
$\alpha\rightarrow -\beta$, $\beta\rightarrow \alpha$.
Again, we can see that $|\text{HN}\rangle_{CS}$ and $|\text{LN}\rangle_{CS}$ have been decomposed into two-meson and hidden-color components.
Examining these expressions reveals the significance of the hidden-color component in both $|\text{LN}\rangle_{CS}$ and $|\text{HN}\rangle_{CS}$.
For the two nonets, percentages of the hidden-color component are as follows:
\begin{eqnarray}
&&\text{For the heavy nonet:}\nonumber \\
&&\left (\frac{\alpha}{\sqrt{6}}+\frac{\beta}{2} \right )^2 + \left (\frac{\alpha}{\sqrt{2}} - \frac{\beta}{2\sqrt{3}}\right )^2 \approx \frac{5}{9}\ ,\\
&&\text{For the light nonet:}\nonumber \\
&&\left (-\frac{\beta}{\sqrt{6}}+\frac{\alpha}{2} \right )^2 + \left (\frac{\beta}{\sqrt{2}} + \frac{\alpha}{2\sqrt{3}}\right )^2 \approx \frac{4}{9}\ .
\end{eqnarray}
This contrasts with the situation before the mixing where the hidden-color component significantly influences $|\text{CS1}\rangle$ ($\approx 2/3$) but becomes less
significant in $|\text{CS2}\rangle$ ($\approx 1/3$).

\section{Hidden-color contribution to hyperfine mass}
\label{hyperfine}

Up to now, we have demonstrated that the tetraquark wave functions, whether before or after the mixing, contain a fair amount of hidden-color components.
As previously discussed, hidden-color components are crucial in distinguishing tetraquarks from meson molecules
because hidden-color components are absent in meson molecules by definition.
Thus, in order to confirm the existence of tetraquarks, it is vital to quantify the contribution from hidden-color components in certain physical quantities.
In this section, we explore the role of hidden-color components in generating the hyperfine mass, represented by $\langle V_{CS}\rangle$.

From Eq.~(\ref{VCS}), the hyperfine mass can be written as
\begin{eqnarray}
\langle V_{CS}\rangle=\sum_{i < j} \frac{v_0}{m_i^{} m_j^{}} \langle V_{ij}\rangle\label{break}\ ,
\end{eqnarray}
where we define $V_{ij}\equiv\lambda_i \cdot \lambda_j J_i\cdot J_j$ for the sake of convenience in our discussion below.
Here, the expectation values can be calculated with respect to various wave functions introduced above.
However, it should be remembered that the physical hyperfine masses correspond to the expectation values
from the physical wave functions in Eqs.~(\ref{heavy}) and (\ref{light}).

To start, we first calculate $\langle V_{ij}\rangle$ with respect to the color-spin wave functions,
$|\text{CS1}\rangle$ and $|\text{CS2}\rangle$ in Eqs.~(\ref{cs1}),(\ref{cs2}).
In Table~\ref{decomposition1}, we present our calculations for $\langle\text{CS1}|V_{ij}|\text{CS1}\rangle$, $\langle\text{CS2}|V_{ij}|\text{CS2}\rangle$,
and $\langle\text{CS1}|V_{ij}|\text{CS2}\rangle$. Each of them comprises six elements depending on $i,j$.
What is new in this calculation is the three contributions
named as TM$\cdot$TM, TM$\cdot$HC, HC$\cdot$HC in Table~\ref{decomposition1}.
These three terms appear in $\langle V_{ij}\rangle$ because
$|\text{CS1}\rangle$ and $|\text{CS2}\rangle$ are divided into the two-meson (TM) part and the hidden-color (HC) part.
The TM$\cdot$TM part represents the contribution solely from the two-meson component, whereas the TM$\cdot$HC and HC$\cdot$HC terms
represent the mixing contribution and pure hidden-color contribution, respectively.
The summation of these three contributions gives $\langle V_{ij}\rangle$ listed in the second column of Table~\ref{decomposition1}. It has been verified that
the values in the second column are consistent with the results presented in table 2 of Ref.~\cite{Kim:2016dfq}.

As shown in Table~\ref{decomposition1}, the hidden-color contributions, through the TM$\cdot$HC, HC$\cdot$HC parts,
are crucial to generate $\langle V_{ij}\rangle$ in most cases
with a few exceptions such as $\langle\text{CS2}|V_{13}|\text{CS2}\rangle$, $\langle\text{CS2}|V_{24}|\text{CS2}\rangle$, $\langle\text{CS1}|V_{13}|\text{CS2}\rangle$, and $\langle\text{CS1}|V_{24}|\text{CS2}\rangle$.
In particular, there is no contribution at all from the TM$\cdot$TM part for all the six members of $\langle\text{CS1}|V_{ij}|\text{CS1}\rangle$.
In other cases as well, the hyperfine masses are primarily given by the TM$\cdot$HC and HC$\cdot$HC contributions.
This clearly shows that hidden-color components are important to generate the hyperfine masses, $\langle V_{ij}\rangle$, calculated in the basis of $|\text{CS1}\rangle$, $|\text{CS2}\rangle$.

\begin{table}[t]
\centering
\begin{tabular}{c|c|ccc}\hline
\multicolumn{2}{l|}{$\langle\text{CS1}|V_{ij}|\text{CS1}\rangle$} & TM$\cdot$TM & TM$\cdot$HC & HC$\cdot$HC \\
\hline
$\langle V_{12}\rangle$& $2$ & $0$ & $\frac{4}{3}$ & $\frac{2}{3}$      \\[1mm]
$\langle V_{13}\rangle$& $0$ & $0$ & $0$ & $0$       \\[1mm]
$\langle V_{14}\rangle$& $0$ & $0$ & $0$ & $0$      \\[1mm]
$\langle V_{23}\rangle$& $0$ & $0$ & $0$ & $0$    \\[1mm]
$\langle V_{24}\rangle$& $0$ & $0$ & $0$ & $0$   \\[1mm]
$\langle V_{34}\rangle$& $2$ & $0$ & $\frac{4}{3}$ & $\frac{2}{3}$  \\[1mm]
\hline
\multicolumn{2}{l|}{$\langle\text{CS2}|V_{ij}|\text{CS2}\rangle$} & TM$\cdot$TM & TM$\cdot$HC & HC$\cdot$HC \\
\hline
$\langle V_{12}\rangle$ & $\frac{1}{3}$& $0$ & $\frac{4}{9}$ & $-\frac{1}{9}$     \\[1mm]
$\langle V_{13}\rangle$ & $\frac{5}{3}$& $\frac{16}{9}$ & $0$ & $-\frac{1}{9}$      \\[1mm]
$\langle V_{14}\rangle$ & $\frac{5}{3}$& $0$ & $\frac{8}{9}$ & $\frac{7}{9}$      \\[1mm]
$\langle V_{23}\rangle$ & $\frac{5}{3}$& $0$ & $\frac{8}{9}$ & $\frac{7}{9}$   \\[1mm]
$\langle V_{24}\rangle$ & $\frac{5}{3}$& $\frac{16}{9}$ & $0$ & $-\frac{1}{9}$   \\[1mm]
$\langle V_{34}\rangle$ & $\frac{1}{3}$& $0$ & $\frac{4}{9}$ & $-\frac{1}{9}$   \\[1mm]
\hline
\multicolumn{2}{l|}{$\langle\text{CS1}|V_{ij}|\text{CS2}\rangle$}  & TM$\cdot$TM & TM$\cdot$HC & HC$\cdot$HC \\
\hline
$\langle V_{12}\rangle$ & $0$ & $0$ & $0$ & $0$ \\[1mm]
$\langle V_{13}\rangle$ & $\sqrt{\frac{3}{2}}$ & $\frac{4\sqrt{6}}{9}$ & $0$ & $\frac{\sqrt{6}}{18}$  \\[1mm]
$\langle V_{14}\rangle$ & $\sqrt{\frac{3}{2}}$ & $0$ & $\frac{\sqrt{6}}{9}$ & $\frac{7\sqrt{6}}{18}$  \\[1mm]
$\langle V_{23}\rangle$ & $\sqrt{\frac{3}{2}}$ & $0$ & $\frac{\sqrt{6}}{9}$ & $\frac{7\sqrt{6}}{18}$  \\[1mm]
$\langle V_{24}\rangle$ & $\sqrt{\frac{3}{2}}$& $\frac{4\sqrt{6}}{9}$ & $0$ & $-\frac{\sqrt{6}}{18}$  \\[1mm]
$\langle V_{34}\rangle$ & $0$ & $0$ & $0$ & $0$  \\[1mm]
\end{tabular}
\caption{Here we present $\langle\text{CS1}|V_{ij}|\text{CS1}\rangle$, $\langle\text{CS2}|V_{ij}|\text{CS2}\rangle$,
and $\langle\text{CS1}|V_{ij}|\text{CS2}\rangle$ in the second column. The other three columns show individual contributions (TM$\cdot$TM, TM$\cdot$HC, HC$\cdot$HC)
that comprise the $\langle V_{ij}\rangle$ given in the second column.}
\label{decomposition1}
\end{table}

The results in Table~\ref{decomposition1} can be directly used
to obtain the hyperfine masses in the mixing wave functions,
$|\text{HN}\rangle_{CS}$ and $|\text{LN}\rangle_{CS}$ in Eqs.~(\ref{hcs}),(\ref{lcs}).
The hyperfine mass in this basis can be calculated as:
\begin{eqnarray}
\langle\text{HN}|V_{ij}|\text{HN}\rangle_{CS}&&=\alpha^2\langle\text{CS1}|V_{ij}|\text{CS1}\rangle -2\alpha\beta \langle\text{CS1}|V_{ij}|\text{CS2}\rangle\nonumber \\
&&+\beta^2\langle\text{CS2}|V_{ij}|\text{CS2}\rangle\label{HNCS} ,\\
\langle\text{LN}|V_{ij}|\text{LN}\rangle_{CS}&&=\beta^2\langle\text{CS1}|V_{ij}|\text{CS1}\rangle +2\alpha\beta \langle\text{CS1}|V_{ij}|\text{CS2}\rangle\nonumber \\
&&+\alpha^2\langle\text{CS2}|V_{ij}|\text{CS2}\rangle\label{LNCS} .
\end{eqnarray}
So, simply plugging the numbers from Table~\ref{decomposition1} in these equations leads to the formulas listed in Table~\ref{decomposition2}.
Note, $\langle\text{LN}|V_{ij}|\text{LN}\rangle_{CS}$ can be obtained alternatively from $\langle\text{HN}|V_{ij}|\text{HN}\rangle_{CS}$ by
replacing $\alpha \rightarrow -\beta$, $\beta \rightarrow \alpha$.
Another thing to note is that,
through Eqs.~(\ref{HNCS}),(\ref{LNCS}), the three parts separately calculated in Table~\ref{decomposition1}
have been transferred into the corresponding
three parts denoted as TM$\cdot$TM, TM$\cdot$HC, and HC$\cdot$HC in Table~\ref{decomposition2}.
Their sum also yields the hyperfine masses listed in the second column of this table.
We see that the HC$\cdot$HC term is always nonzero for all the elements while the TM$\cdot$TM term is nonzero only for certain elements.
Hence, the hidden-color contribution should be important in generating the hyperfine masses, $\langle V_{ij}\rangle$, calculated in the mixing
wave functions.

The results in Table~\ref{decomposition2} can be utilized to compute the physical hyperfine masses
in the basis of $|\text{Heavy nonet}\rangle$ and $|\text{Light nonet}\rangle$ [Eqs.~(\ref{heavy}) and (\ref{light})].
These physical states possess additional flavor structures,
given by Eqs.~(\ref{flavor octet}) (\ref{flavor singlet}),
that must be incorporated through the factor $\frac{v_0}{m_im_j}$ in $\langle V_{CS}\rangle$ of Eq.~(\ref{break}).
For instance, the hyperfine mass for a specific flavor structure represented by $q_1q_2\bar{q}_3\bar{q}_4$ can be computed by multiplying
each $\langle V_{ij}\rangle$ with its corresponding $\frac{v_0}{m_im_j}$ and summing over all quark pairs.
Mathematically, this hyperfine mass can be expressed as:
\begin{eqnarray}
\langle\text{HN}|V_{CS}|\text{HN}\rangle_{q_1q_2\bar{q}_3\bar{q}_4}\!\! &=& \!\! \sum_{i>j}\frac{v_0}{m_im_j}\langle\text{HN}|V_{ij}|\text{HN}\rangle_{CS}\label{HNf}\ ,\\
\langle\text{LN}|V_{CS}|\text{LN}\rangle_{q_1q_2\bar{q}_3\bar{q}_4}\!\! &=& \!\! \sum_{i>j}\frac{v_0}{m_im_j}\langle\text{LN}|V_{ij}|\text{LN}\rangle_{CS}\label{LNf}\ .
\end{eqnarray}

\begin{table}[t]
\centering
\begin{tabular}{c|c|c|c|c}\hline
\multicolumn{2}{c|}{$\langle\text{HN}|V_{ij}|\text{HN}\rangle_{CS}$} & TM$\cdot$TM & TM$\cdot$HC & HC$\cdot$HC \\
\hline
$\langle V_{12}\rangle$ & $2\alpha^2+\frac{\beta^2}{3}$ & $0$ & $\frac{12\alpha^2+4\beta^2}{9}$ & $\frac{6\alpha^2-\beta^2}{9}$  \\[1mm]
$\langle V_{13}\rangle$ & $-\sqrt{6}\alpha\beta+\frac{5}{3}\beta^2$ & $\frac{-8\sqrt{6}\alpha\beta+16\beta^2}{9}$ & $0$ & $\frac{-\sqrt{6}\alpha\beta-\beta^2}{9}$      \\[1mm]
$\langle V_{14}\rangle$ & $-\sqrt{6}\alpha\beta+\frac{5}{3}\beta^2$ & $0$ & $\frac{-2\sqrt{6}\alpha\beta+8\beta^2}{9}$ & $\frac{-7\sqrt{6}\alpha\beta+7\beta^2}{9}$     \\[1mm]
$\langle V_{23}\rangle$ & $-\sqrt{6}\alpha\beta+\frac{5}{3}\beta^2$ & $0$ & $\frac{-2\sqrt{6}\alpha\beta+8\beta^2}{9}$ & $\frac{-7\sqrt{6}\alpha\beta+7\beta^2}{9}$      \\[1mm]
$\langle V_{24}\rangle$ & $-\sqrt{6}\alpha\beta+\frac{5}{3}\beta^2$ & $\frac{-8\sqrt{6}\alpha\beta+16\beta^2}{9}$ & $0$ & $\frac{-\sqrt{6}\alpha\beta-\beta^2}{9}$      \\[1mm]
$\langle V_{34}\rangle$ & $2\alpha^2+\frac{\beta^2}{3}$ & $0$ & $\frac{12\alpha^2+4\beta^2}{9}$ & $\frac{6\alpha^2-\beta^2}{9}$   \\[1mm]
\hline
\multicolumn{2}{c|}{$\langle\text{LN}|V_{ij}|\text{LN}\rangle_{CS}$} & TM$\cdot$TM & TM$\cdot$HC & HC$\cdot$HC\\
\hline
$\langle V_{12}\rangle$ & $2\beta^2+\frac{\alpha^2}{3}$ & $0$ & $\frac{12\beta^2+4\alpha^2}{9}$ & $\frac{6\beta^2-\alpha^2}{9}$    \\[1mm]
$\langle V_{13}\rangle$ & $\sqrt{6}\alpha\beta+\frac{5}{3}\alpha^2$ & $\frac{8\sqrt{6}\alpha\beta+16\alpha^2}{9}$ & $0$ & $\frac{\sqrt{6}\alpha\beta-\alpha^2}{9}$      \\[1mm]
$\langle V_{14}\rangle$ & $\sqrt{6}\alpha\beta+\frac{5}{3}\alpha^2$ & $0$ & $\frac{2\sqrt{6}\alpha\beta+8\alpha^2}{9}$ & $\frac{7\sqrt{6}\alpha\beta+7\alpha^2}{9}$      \\[1mm]
$\langle V_{23}\rangle$ & $\sqrt{6}\alpha\beta+\frac{5}{3}\alpha^2$ & $0$ & $\frac{2\sqrt{6}\alpha\beta+8\alpha^2}{9}$ & $\frac{7\sqrt{6}\alpha\beta+7\alpha^2}{9}$     \\[1mm]
$\langle V_{24}\rangle$ & $\sqrt{6}\alpha\beta+\frac{5}{3}\alpha^2$ & $\frac{8\sqrt{6}\alpha\beta+16\alpha^2}{9}$ & $0$ & $\frac{\sqrt{6}\alpha\beta-\alpha^2}{9}$       \\[1mm]
$\langle V_{34}\rangle$ & $2\beta^2+\frac{\alpha^2}{3}$ & $0$ & $\frac{12\beta^2+4\alpha^2}{9}$ & $\frac{6\beta^2-\alpha^2}{9}$     \\[1mm]
\end{tabular}
\caption{Here, we present the expectation values of $\langle\text{HN}|V_{ij}|\text{HN}\rangle_{CS}$ and $\langle\text{LN}|V_{ij}|\text{LN}\rangle_{CS}$,
which are deduced from Eqs.~(\ref{HNCS}),(\ref{LNCS}).
As in Table~\ref{decomposition1}, we also present separate contribution from TM$\cdot$TM, TM$\cdot$HC, and HC$\cdot$HC so we can see
how the results in the second column are broken down into these three contributions.
}
\label{decomposition2}
\end{table}

But, as evident from Eq.~(\ref{flavor octet}) or Eq.~(\ref{flavor singlet}), the flavor wave function is not
restricted to a single combination but encompasses various combinations that must also be taken into account.
To exemplify this, we consider the isovector resonance, $a_0^+(1450)$, in the heavy nonet, which has the flavor wave function
described by Eq.~(\ref{o21}):
\begin{eqnarray}
[su][\bar{d}\bar{s}]=\frac{1}{2}\big\{su\bar{d}\bar{s}-su\bar{s}\bar{d}-us\bar{d}\bar{s}+us\bar{s}\bar{d}\big\}\label{isovector1}\ .
\end{eqnarray}
The physical hyperfine mass for $a_0^+(1450)$, therefore, can be calculated as
\begin{eqnarray}
\langle V_{CS}\rangle &&=\frac{1}{4}\Big\{\langle\text{HN}|V_{CS}|\text{HN}\rangle_{su\bar{d}\bar{s}}+\langle\text{HN}|V_{CS}|\text{HN}\rangle_{su\bar{s}\bar{d}}\nonumber \\
&&+\langle\text{HN}|V_{CS}|\text{HN}\rangle_{us\bar{d}\bar{s}}+\langle\text{HN}|V_{CS}|\text{HN}\rangle_{us\bar{s}\bar{d}}\Big\}\label{isovectorh}\ ,
\end{eqnarray}
where each term in the right-hand side is evaluated according to Eq.~(\ref{HNf}) with the specified flavor in subscripts.
For the corresponding member in the light nonet, $a_0^+(980)$,  its hyperfine mass can be calculated similarly:
\begin{eqnarray}
\langle V_{CS}\rangle &&=\frac{1}{4}\Big\{\langle\text{LN}|V_{CS}|\text{LN}\rangle_{su\bar{d}\bar{s}}+\langle\text{LN}|V_{CS}|\text{LN}\rangle_{su\bar{s}\bar{d}}\nonumber \\
&&+\langle\text{LN}|V_{CS}|\text{LN}\rangle_{us\bar{d}\bar{s}}+\langle\text{LN}|V_{CS}|\text{LN}\rangle_{us\bar{s}\bar{d}}\Big\}\label{isovectorl}\ ,
\end{eqnarray}
where each term in the right-hand side is evaluated according to Eq.~(\ref{LNf}).
Since the major ingredients, $\langle\text{HN}|V_{ij}|\text{HN}\rangle_{CS}$ and $\langle\text{LN}|V_{ij}|\text{LN}\rangle_{CS}$,
have been calculated separately for the three parts in Table~\ref{decomposition2},
the hyperfine masses in the physical basis can be computed separately also for the TM$\cdot$TM, TM$\cdot$HC, HC$\cdot$HC contributions.

The physical hyperfine masses, calculated from formulas similar to Eqs.~(\ref{isovectorh}) and (\ref{isovectorl}),
provide successful phenomenology of the tetraquark mixing model. Based on the analysis above, it is quite likely that hidden-color components play
a significant role in generating the physical hyperfine mass. Given the uniqueness of hidden-color components in tetraquarks,
demonstrating their contribution could provide direct evidence for the existence of tetraquarks.
Below we present successful aspects of the hyperfine masses in the tetraquark mixing model and discuss the
contributions from the hidden-color components.

Table~\ref{hyperfine1} shows the physical hyperfine masses for the isovector and isodoublet members in the heavy and light nonets.
The input parameters for this calculation are $v_0=(-192.9)^3$ MeV$^3$, $m_u=m_d=330$ MeV, and $m_s=500$ MeV,
which are the same values used in Refs.~\cite{Kim:2016dfq, Kim:2017yvd, Kim:2018zob}.
Here, we present the separate contributions from TM$\cdot$TM, TM$\cdot$HC, HC$\cdot$HC and
their sum that yields the hyperfine masses in the second column of Table~\ref{hyperfine1}.
Table~\ref{hyperfine2} shows the hyperfine masses for the isoscalar resonances. In the isoscalar channel, we present
three cases depending on how the flavor mixing is implemented~\cite{Kim:2018zob}.
To validate our calculations, we have ensured that the hyperfine masses in the second column of Table~\ref{hyperfine1} and Table~\ref{hyperfine2}
indeed match the previous results reported in Refs.~\cite{Kim:2016dfq,Kim:2017yvd,Kim:2018zob}.

\begin{table}[t]
\centering
\begin{tabular}{c|r|rrr}\hline
Resonance & $\langle V_{CS}\rangle$ & TM$\cdot$TM & TM$\cdot$HC & HC$\cdot$HC   \\
\hline
$a_0(980)$  & $-489$ & $-201$ & $-141$ & $-146$       \\[1mm]
$K_0^*(700)$ & $-593$ & $-241$ & $-174$ & $-178$      \\[1mm]
$a_0(1450)$ & $-17$ & $39$ & $-94$ & $38$     \\[1mm]
$K_0^*(1430)$ & $-27$& $47$ & $-118$ & $44$       \\
\hline
\end{tabular}
\caption{The hyperfine masses for the isosvector and isodoublet members in both the light and heavy nonets are presented here in MeV units.
Also the three individual contributions that comprise the hyperfine mass given in the second column.
 }
\label{hyperfine1}
\end{table}

\begin{table}[t]
\centering
\begin{tabular}{c|r|rrr}\hline
Resonance & $\langle V_{CS}\rangle$ & TM$\cdot$TM & TM$\cdot$HC & HC$\cdot$HC  \\
\hline
\multicolumn{1}{r}{\underline{Case1}} \\
$f_0(500)$ & $-668$ & $-272$ & $-196$ & $-200$       \\[1mm]
$f_0(980)$ & $-535$ & $-219$ & $-156$ & $-160$    \\[1mm]
$f_0(1370)$ & $-29$ & $53$ & $-132$ & $50$      \\[1mm]
$f_0(1500)$ & $-20$ & $43$ & $-104$ & $41$      \\[1mm]
\hline
\multicolumn{1}{r}{\underline{Case2}} \\
$f_0(500)$  & $-639$  & $-261$ & $-187$ & $-191$     \\[1mm]
$f_0(980)$  & $-564$ & $-231$ & $-164$ & $-169$    \\[1mm]
$f_0(1370)$ & $-27$ & $51$ & $-126$ & $48$      \\[1mm]
$f_0(1500)$ & $-22$ & $45$ & $-110$ & $43$     \\[1mm]
\hline
\multicolumn{1}{r}{\underline{Case3}} \\
$f_0(500)$  & $-714$ & $-291$ & $-210$ & $-214$     \\[1mm]
$f_0(980)$  & $-489$ & $-201$ & $-141$ & $-146$   \\[1mm]
$f_0(1370)$ & $-32$ & $56$ & $-142$ & $53$      \\[1mm]
$f_0(1500)$ & $-17$ & $39$ & $-94$ & $38$     \\[1mm]
\hline
\end{tabular}
\caption{The hyperfine masses for the isoscalar members in both the light and heavy nonets are presented here in MeV units.
For isoscalar resonances, we present three different cases depending on how the flavor mixing is implemented as discussed in Ref.~\cite{Kim:2018zob}.
Namely, Case1 is the realistic case with fitting, Case2 is the SU(3) symmetric case, and Case3 is the ideal mixing case.
In each line, we display three individual contributions that constitute the hyperfine mass in the second column.
 }
\label{hyperfine2}
\end{table}

Our conclusion from Table~\ref{hyperfine1} and Table~\ref{hyperfine2} is that the hidden-color component
contributes significantly to the physical hyperfine mass for the light nonet and the heavy nonet as well.
For the $a_0(980)$ in the light nonet, as shown in Table~\ref{hyperfine1}, its hyperfine mass is
$-489$ MeV, large enough to lower the $a_0(980)$ mass below 1 GeV.
A similar trend can be seen in the other members of the light nonet.
This is one of the successful aspects of the tetraquark mixing model because
it can provide a qualitative explanation for why the light nonet, despite consisting of four constituent quarks, has masses below 1 GeV.
However, from Table~\ref{hyperfine1}, we can see that the contribution from the TM$\cdot$TM part is only $-201$ MeV.
The remaining contribution comes from the hidden-color component
through TM$\cdot$HC, HC$\cdot$HC, which respectively contributes $-141$, $-146$ MeV.
Thus, without the hidden-color component, the hyperfine mass cannot sufficiently lower the $a_0(980)$ mass below 1 GeV.

\begin{table}[t]
\centering
\begin{tabular}{c|c||c|c|c}\hline
  Resonance &  $\Delta \langle V_{CS} \rangle$ & TM$\cdot$TM & TM$\cdot$HC & HC$\cdot$HC \\
\hline
 $a_0(1450)-a_0(980)$   &  $472$ & $240$ & $48$     & $184$    \\
 $K_0^*(1430)-K_0^*(700)$ & $566$ & $288$ & $56$     & $222$    \\
 $f_0(1370)-f_0(500)$   &  $612$ & $311$ & $61$     & $239$    \\
 $f_0(1500)-f_0(980)$  &  $542$ & $276$ & $54$     & $212$    \\
\hline
\end{tabular}
\caption{Here we present the hyperfine mass splitting, $\Delta \langle V_{CS} \rangle$, for each isospin channel, and
three different contributions from TM$\cdot$TM, TM$\cdot$HC and HC$\cdot$HC.
For the isoscalar resonances, the results here show only the Case2 in Table~\ref{hyperfine2} as the other cases lead to
the similar conclusion.
 }
\label{hyperfine3}
\end{table}

In the case of the heavy nonet, the hyperfine mass is found to be small around $-20$ MeV or so.
This is another successful aspect of the tetraquark mixing model, as it can explain qualitatively
why the masses of the heavy nonet do not significantly deviate from four times the constituent quark mass.
This small hyperfine mass can be understood also from large contribution from the hidden-color component.
For example, the $a_0(1450)$ has the small hyperfine mass, $-17$ MeV, as shown in Table~\ref{hyperfine1}.
It can be seen that this small value is a consequence of a strong cancelation among the three
contributions since TM$\cdot$TM$=39$ MeV, TM$\cdot$HC$=-94$ MeV, and HC$\cdot$HC$=38$ MeV. This indicates that
the hidden-color contributions are important for generating the small hyperfine mass for the heavy nonet.
We can do a similar analysis for the other resonances in the two nonets.
While the numerical values vary depending on the specific resonances, we have the similar conclusion from the other members.

Another successful aspect of the tetraquark mixing model is its ability to explain
the mass difference between the two nonets through the hyperfine mass splitting, as described by the formula~\cite{Kim:2016dfq,Kim:2017yvd}:
\begin{eqnarray}
\Delta M \approx \Delta \langle V_{CS}\rangle\label{splitting}\ .
\end{eqnarray}
This formula is similar to the one used for mass splitting among the lowest-lying mesons and baryons~\cite{Kim:2014ywa,Lipkin:1986dx},
Thus, it is nice to see that the tetraquark mixing model provides a similar mass splitting formula for the two nonets.
Once again, the hidden-color component is found to substantially contribute to $\Delta\langle V_{CS}\rangle$.
Table~\ref{hyperfine3} shows the hyperfine mass splitting, $\Delta \langle V_{CS}\rangle$, in each isospin channel, and three different contributions
that make the hyperfine splitting.
Through the TM$\cdot$HC and HC$\cdot$HC parts, the hidden-color component contributes approximately half to $\Delta \langle V_{CS}\rangle$.
Therefore, the hidden-color component is indispensable, because without its contribution,
achieving the mass splitting formula like Eq.~(\ref{splitting}) for the two nonets would not be feasible.

As mentioned, hidden-color components distinguish multiquarks such as tetraquarks from hadronic molecules.
Our results demonstrate that the hidden-color contribution
is crucial in generating the hyperfine mass and provides
additional evidence supporting the tetraquark nature of the two nonets.

\section{Summary}
\label{sec:summary}

In this work, we discuss that multiquarks such as tetraquarks can be distinguished
from hadronic molecules by the presence of hidden-color components.
We therefore investigate hidden-color components particularly in the tetraquark mixing wave functions
that have been suggested as a relevant structure for describing the two nonets:
the heavy nonet [$a_0 (1450)$, $K_0^* (1430)$, $f_0 (1370)$, $f_0 (1500)$] and
the light nonet [$a_0 (980)$, $K_0^* (700)$, $f_0 (500)$, $f_0 (980)$].
Specifically, we examine hidden-color contributions to the hyperfine masses of the two nonets.

In the tetraquark mixing model, the wave functions of the two nonets are expressed by linear combinations of two types of tetraquarks, $|\text{Type1}\rangle$ and $|\text{Type2}\rangle$, constructed in the diquark-antidiquark form.
In order to quantify the contribution from hidden-color component, we separate the two tetraquark types
into two-meson parts and hidden-color parts in color and spin configurations.
Plugging these two types into the tetraquark mixing model, we also decompose the mixing wave functions for the two nonets
into two-meson parts and hidden-color parts.
Hidden-color components are found to constitute a fairly large part, even comparable to two-meson components, in the tetraquark mixing wave functions.
This decomposition allows us to calculate the contribution from hidden-color components to the hyperfine mass of the two nonets.

Our findings reveal a significant contribution from the hidden-color components to the physical hyperfine masses of both the light nonet and the heavy nonet.
In the light nonet, the hyperfine mass is quite large, approximately $-500$ MeV, with the hidden-color components contributing roughly 50 percent.
Conversely, in the heavy nonet, the hyperfine mass is as small as around $-20$ MeV.
However, this value is also a consequence of a large contribution from the hidden-color components
as it is from strong cancelation among those from the two-meson components and the hidden-color components.
Given that the hidden-color components are unique to multiquarks such as tetraquarks, the presence of their contribution in the hyperfine mass provides
additional evidence supporting the tetraquark nature of the two nonets.

\acknowledgments
This work was supported by the National Research Foundation of Korea(NRF) grant funded by the
Korea government(MSIT) (No. NRF-2023R1A2C1002541, No. NRF-2018R1A5A1025563).

\end{document}